\documentclass[twocolumn, prl, longbibliography]{revtex4-2}
\usepackage{amsmath}
\usepackage{amsbsy}
\usepackage{amssymb}
\usepackage{graphicx}
\usepackage{color}
\usepackage{xcolor}
\usepackage{subfigure}
\usepackage{physics}
\usepackage{soul}
\usepackage{color}
\usepackage{bm}
\usepackage{array}
\usepackage{multirow}
\usepackage{lipsum}
\usepackage[normalem]{ulem}
\usepackage{verbatim}
\usepackage{natbib}
\usepackage{nccmath}
\usepackage[pdftex,colorlinks=true,
	pdfstartview = FitV,
	linkcolor    = linkcolor,
	citecolor    = linkcolor,
	urlcolor     = linkcolor,	
	hyperindex   = true,
	hyperfigures = false]{hyperref}
	
\definecolor{linkcolor}{rgb}{0,0,0.6}

\usepackage{cleveref}
	
\usepackage{lineno}

\begin{document} 

\title{Pulsating active matter}

\author{Yiwei Zhang}
\author{\'Etienne Fodor}
\affiliation{Department of Physics and Materials Science, University of Luxembourg, L-1511 Luxembourg, Luxembourg}

\begin{abstract}
We reveal that the mechanical pulsation of locally synchronised particles is a generic route to propagate deformation waves. We consider a model of dense repulsive particles whose activity drives periodic change in size of each individual. The dynamics is inspired by biological tissues where cells consume fuel to sustain active deformation. We show that the competition between repulsion and synchronisation triggers an instability which promotes a wealth of dynamical patterns, ranging from spiral waves to defect turbulence. We identify the mechanisms underlying the emergence of patterns, and characterize the corresponding transitions. By coarse-graining the dynamics, we propose a hydrodynamic description of an assembly of pulsating particles, and discuss an analogy with reaction-diffusion systems.
\end{abstract}

\maketitle
%\linenumbers

Active matter features injection of energy at the individual level to sustain nonequilibrium dynamics~\cite{Marchetti2013, Wijland2022, Jack2022}. The interplay between particle interaction and individual activity opens the door to collective behaviours without any equilibrium counterpart. In many studies, activity takes the form of self-propulsion: each particle converts the energy provided by some fuel into translational motion. Depending on microscopic symmetries, self-propelled particles can exhibit flocking transition~\cite{Chate2020} and/or motility-induced phase separation~\cite{Cates2015}, for instance. In dense regimes, self-propulsion shifts the glass transition~\cite{Ni2013, Berthier2019}. It also controls the solid-fluid transition in models of cellular tissues, where the area and perimeter of each cell are constrained~\cite{Bi2015, Bi2016}.

Activity is not restricted to self-propulsion in general. In biological tissues, living cells use chemical energy to undergo mechanical deformation, and also to power division, extrusion, nematic stresses. For instance, cells can collectively increase and decrease their sizes in a sustained and locally synchronised fashion~\cite{Zehnder2015, Thiagarajan2022}. To maintain close packing at constant area, without any free boundary, local increase in density induces nearby decrease: tissues must accommodate coexistence between contracting and expanding regions. This promotes propagation of contraction waves and pulses, as reported both in vivo~\cite{Martin2009, Solon2009, Armon2018} and in vitro~\cite{Serra-picamal2012, Zehnder2015, Thiagarajan2022, Tlili2018, Petrolli2019, Mueller2019, Hino2020, Boocock2021}.

The emergence of collective contraction is of tremendous importance in many biological contexts. During morphogenesis, mechanochemical coupling yields wave propagation driving the early stages of embryonic development~\cite{Heisenberg2013, Lecuit2019, Lecuit2022}. In cardiac tissue, electromechanical coupling leads to contraction pulses spontaneously organizing in patterns~\cite{Karma1993, Luther2018, Karma2022}, some of which signal arrhythmogenesis~\cite{Karma2013, Rappel2022}. In uterine tissue, a similar coupling regulates large-scale contraction during labor~\cite{Pumir2015, Elad2017}. For all cases, understanding how to control contractile patterns is a first step towards mitigating various health issues.

Some models, from particles~\cite{Petrolli2019, Armon2021} to hydrodynamics~\cite{Salbreux2014, Banerjee2015}, capture contraction waves by regarding activity as a combination of self-propulsion and contraction. Yet, it can be surprising to regard translational motion as the dominant nonequilibrium factor when cells barely move. Indeed, even when biological tissues behave as solids, collective states supported by individual deformation can still emerge. Some preliminary works have considered models where activity sustains size oscillations~\cite{Tjhung2016, Tjhung2017, Togashi2019, Koyano2019, Oyama2019}, which indeed leads to novel physics compared with that of self-propelled particles.

\begin{figure}[b]
  \includegraphics[width=.9\linewidth, trim=0 0 0 0, clip=true]{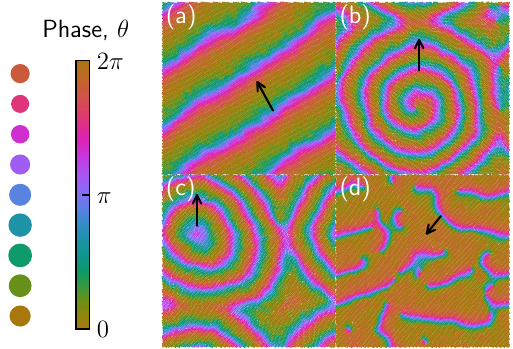}
  \caption{Repulsive particles with pulsating size yield deformation waves: (a)~planar, (b)~spiral, (c)~circular, and (d)~turbulent. Wave propagation (black arrows) stabilizes dynamical patterns reminiscent of reaction-diffusion systems~\cite{movies}.
  }
	\label{Fig:0}
\end{figure}

In this Letter, we formulate a model of pulsating active matter (PAM) which reveals that mechanical pulsation is key to a wealth of dynamical phenomena [Fig.~\ref{Fig:0}]\cite{movies}. We characterize the transitions associated with the emergence of deformation waves, and show that the corresponding hydrodynamics bear analogy to reaction-diffusion systems (RDS)~\cite{Turing1952, Kondo2010}. Although our approach is inspired by the specific case of biological tissues, our model should be regarded as a minimal, yet non-trivial description of a broader class of systems made of active deforming particles. In that respect, our results unveil a generic mechanism for propagating deformation waves in a dense environment: the competition between repulsion and synchronisation under mechanical pulsation.

% ------------------------------------------------------------------------------------------

\textit{Model.}---We consider $N$ deforming particles in two dimensions subject to the pairwise repulsive potential $U$. The state of each particle is described by its phase $\theta$, determining the particle size, and position $\mathbf{r}$. The overdamped dynamics of positions $\{{\bf r}_i\}$ reads
\begin{equation}\label{eqn:MD}
	\dot{\bf r}_i = -\mu \sum_j \partial_{{\bf r}_i} U (a_{ij}) + \sqrt{2D}\, {\boldsymbol\xi}_i , \quad a_{ij} = \frac{|{\bf r}_i - {\bf r}_j|}{\sigma(\theta_i)+\sigma(\theta_j)} ,
\end{equation}
where $\mu$ and $D$ are respectively the mobility and diffusivity, and ${\boldsymbol\xi}_i$ is an isotropic Gaussian white noise with zero mean. We take $U(a) = U_0 (a^{-12} - 2 a^{-6})$ for $a<1$, and $U(a)=0$ otherwise. The interaction range is controlled by the particle size $\sigma(\theta_i)$ given by
\begin{equation}
	\sigma(\theta_i) = \sigma_0 \frac{1+\lambda \sin\theta_i}{1+\lambda} ,
\end{equation}
where $\sigma_0$ is the largest size, and $\lambda<1$ describes the pulsation amplitude. We regard the phases $\{\theta_i\}$ as stochastic degrees of freedom subject to the drive $\omega$, which sustains periodic deformation of particles, and to the local synchronisation $\cal T$:
\begin{equation}\label{eqn:MDb}
	\dot{\theta}_i = \omega - \sum_j \Big[ {\cal T}(a_{ij}, \theta_i-\theta_j) + \mu_{\theta} \partial_{\theta_i} U (a_{ij})\Big] + \sqrt{2D_{\theta}} \eta_i ,
\end{equation}
where $\mu_\theta$ and $D_\theta$ are the effective mobility and diffusivity of phases, and $\eta_i$ is a Gaussian white noise with zero mean. The synchronisation range is the same as that of repulsion: ${\cal T} (a,\theta) = \varepsilon\sin(\theta)$ for $a<1$, and ${\cal T} (a,\theta)=0$ otherwise. In the limit of passive deforming particles (namely, for $\omega=0$, $D/\mu=D_\theta/\mu_\theta$, and ${\cal T}=0$), the system can display glassy behavior at high density for an appropriate choice of interaction $U$~\cite{Berthier2017, Wyart2018}.

% ------------------------------------------------------------------------------------------

\textit{Patterns and phases.}---The repulsion terms $\partial_{r_i} U$ and $\partial_{\theta_i} U$ promote a uniform density profile by minimizing particle overlap. Besides, these terms impede the expansion of particles, and facilitate their contraction, whenever they are in contact. In contrast to previous works on driven deforming particles~\cite{Tjhung2016, Tjhung2017}, the packing fraction $\varphi$ here varies in time when particle sizes $\sigma(\theta_i)$ are cycling:
\begin{equation}
	\varphi = \pi \sum_{i=1}^N (\sigma(\theta_i)/L)^2 ,
\end{equation}
where $L$ is the system size. Above a given density, repulsion precludes that all particles simultaneously reach their largest size. Yet, synchronisation puts a cost on phase differences between neighbours, thus favoring a uniform phase profile. The competition between repulsion and synchronisation can then destabilize the uniform phase profile, if the system is large enough to accomodate a finite-wavelength instability, while the density profile stays roughly homogeneous.

This instability promotes deformation waves, propagating in the direction of the drive $\omega$ [Eq.~\eqref{eqn:MDb}], which spontaneously organise into dynamical patterns. Such patterns are either steady, in the form of planar, spiral, or circular waves [Figs.~\ref{Fig:0}(a-c)], or continuously changing [Figs.~\ref{Fig:0}(d)], and some patterns entail defects in the phase profile [Figs.~\ref{Fig:0}(b-d)]. Interestingly, our model features a phenomenology reminiscent of RDS. Yet, in contrast to RDS, the instability underlying pattern formation does not rely here on any chemical reaction, it only stems from mechanical pulsation, repulsion, and synchronisation.

\begin{figure}[b]
  \includegraphics[width=\linewidth]{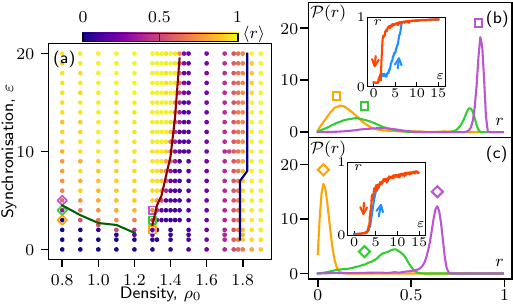}
  \caption{Phases and transitions in particle-based dynamics.
  (a)~We distinguish two ordered states ($\langle r\rangle\approx 1$) where particles are either cycling in phase (small $\rho_0$ and large $\varepsilon$) or arrested (large $\rho_0$). The solid lines in dark green ($\varepsilon_{b,1}(\rho_0)$), dark red ($\varepsilon_{b,2}(\rho_0)$), and dark blue ($\varepsilon_{b,3}(\rho_0)$) are guidelines delineating the boundaries between ordered and disordered states~\cite{SM}. Dynamical patterns and waves [Fig.~\ref{Fig:0}] emerge in the disordered state at large $\varepsilon$, in between the phase boundaries $\varepsilon_{b,2}(\rho_0)$ and $\varepsilon_{b,3}(\rho_0)$.
   (b)~At $\rho_0=1.3$, the distribution ${\cal P}(r)$ changes between unimodal and bimodal shapes when varying $\varepsilon$ through the phase boundary $\varepsilon_{b,2}(\rho_0)$ [squares in (a)]. The transition is discontinuous with hysteresis [inset].
   (c)~At $\rho_0=0.8$, ${\cal P}(r)$ stays unimodal through the boundary $\varepsilon_{b,1}(\rho_0)$ [diamonds in (a)]. The transition is continuous without hysteresis [inset].
  }
	\label{Fig:1}
\end{figure}

We characterize the emergent phenomenology in terms of the phase dynamics, given that the signature of patterns in the position dynamics is less straightforward. The uniform phase profile is a fully synchronized state, with most particles sharing the same phase at each time, whereas deformation waves and defects are associated with phase inhomogeneities. In that respect, we regard the emergence of waves as a transition from order to disorder in terms of the synchronisation parameter
\begin{equation}\label{order_param}
	r = \frac{1}{N}\Bigg| \sum_{j=1}^N e^{{\rm i}\theta_j} \Bigg| .
\end{equation}
Varying the total density $\rho_0$ and the synchronisation strength $\varepsilon$, which regulate the repulsion-synchronisation trade-off, we observe two disconnected regions where the averaged parameter $\langle r\rangle$ is close to unity [Fig.~\ref{Fig:1}(a)]. The ordered state is cycling at small density ($\rho_0<1.5$) and arrested at large density ($\rho_0>1.8$). Waves emerge in between these states (large $\varepsilon$ and moderate $\rho_0$) through a transition destabilizing either the cycling or the arrested ordered state, yielding $\langle r\rangle$ smaller than unity. We regard waves as a locally ordered, yet globally disordered state, at variance with disorder at smaller $\varepsilon$ where the system is both locally \textit{and} globally disordered.

Overall, we delineate three boundaries in the phase diagram~\cite{SM}[Fig.~\ref{Fig:1}(a)]: (i)~at small density ($\rho_0<\rho_c$, $\rho_c\approx 1.3$), $\varepsilon_{b,1}(\rho_0)$ decreases with $\rho_0$ [dark green line], (ii)~at moderate density ($\rho_c<\rho_0<1.5$), $\varepsilon_{b,2}(\rho_0)$ increases with $\rho_0$ [dark red line], and (iii)~at high density ($\rho_0>1.8$), $\varepsilon_{b,3}(\rho_0)$ merely depends on $\varepsilon$ [dark blue line]. In what follows, we analyze the nature of these transitions, identify the corresponding microscopic mechanisms, and construct a hydrodynamic description of PAM revealing an analogy with RDS.

% ------------------------------------------------------------------------------------------

\textit{Ordered states: Cycling and arrest.}---At small density ($\rho_0<\rho_c$), particles barely overlap, so that the effect of repulsion in Eq.~\eqref{eqn:MDb} is subdominant compared with synchronisation. Our setting is then akin to the seminal Kuramoto model~\cite{Ritort2005}, albeit neighbour identities now vary in time. Given that all phases are driven at the same frequency, the system orders whenever the synchronisation overcomes the noise. At mean-field level, this requires $\rho_0 \varepsilon$ greater than a factor proportional to $D_\theta$, which is consistent with $\varepsilon_{b,1}(\rho_0)$ decreasing with $\rho_0$.

At high density ($\rho_0>1.8$), particles are too packed to cycle their phase. The ordered state is arrested, with particle sizes fluctuating around an average value given by close packing, without any phase current. To build intuition on how arrest emerges, we approximate the repulsive term in Eq.~\eqref{eqn:MDb} as $\partial_{\theta_i} U = (\partial_\varphi U) (\partial_{\theta_i}\varphi)$, with $\partial_\varphi U$ independent of $\{\theta_i\}$ to first approximation. Hence, we map the pairwise potential $U$ into the effective one-body potential $U_{\rm eff} = (\partial_\varphi U) \,\varphi$, which is a periodic function of each phase $\theta_i$. The drive tilts this potential as $U_{\rm eff} - \omega\sum_i\theta_i$. At small $\omega$ (equivalently, large $\partial_\varphi U$), phases are trapped in a local minimum of the tilted potential, yielding arrest. In this regime, the potential contribution is overwhelmingly dominant compared with synchronisation. As a result, the phase boundary $\varepsilon_{b,3}(\rho_0)$ weakly depends on $\varepsilon$, and we expect that $\varepsilon_{b,3}(\rho_0)$ is strongly affected by $U_0$ and $\mu_\theta$.

% ------------------------------------------------------------------------------------------

\textit{Beyond ordered states: Waves and defects.}---Reducing $\rho_0$ decreases $\partial_\varphi U$, which lowers the depth of local minima in the tilted potential. Fluctuations can then trigger jumps between minima, which promotes transient cycling of size. As a result, the system now features spatial coexistence between cycling and arrested particles, which yields $\langle r\rangle$ smaller than $1$, and destabilizes the homogeneous phase profile. At large $\varepsilon$, the synchronisation leads local cycling to propagate between nearest neighbours, yielding deformation waves in an arrested background [Fig.~\ref{Fig:0}(d)]. Interestingly, defects are present at both ends of each wave, rotating in opposite directions~\cite{movies}. Such defect pairs spontaneously form, move, and annihilate, thus forming a dynamical state analogous to the defect turbulence reported in RDS~\cite{Ouyang1996}.

As $\rho_0$ gets smaller, more particles are prone to cycling, which reduces the proportion of arrested particles, and yields smaller $\langle r\rangle$. For $\langle r\rangle$ close to $0$, namely when all particles cycle, dynamical patterns spontaneously organize into specific structures: either planar, spiral, or circular waves [Figs.~\ref{Fig:0}(a-c)]. Correspondingly, we observe that the number of defects is much reduced, and defects are also less motile than in defect turbulence. Eventually, further decreasing $\rho_0$, the system enters the cycling ordered state [top left of Fig.~\ref{Fig:1}(a)].

We now examine how the distribution ${\cal P}(r)$ varies through the phase boundaries $\varepsilon_{b,1}(\rho_0)$ and $\varepsilon_{b,2}(\rho_0)$, which respectively decrease and increase with $\rho_0$. For the boundary $\varepsilon_{b,2}(\rho_0)$, ${\cal P}$ is peaked at small $r$ in the disordered state, and it becomes bimodal as the system orders, with weight shifting from small to large $r$ [Fig.~\ref{Fig:1}(b)]. This indicates the existence of a metastable regime where both states are linearly stable. Starting deep in the ordered state and reducing $\varepsilon$, a fluctuation eventually destabilizes the global synchronisation, and triggers a discontinuous transition towards small $r$ [inset of Fig.~\ref{Fig:1}(b)]. Conversely, starting from the disordered state and increasing $\varepsilon$ reveals hysteresis, another signature of metastability. A similar analysis for the boundary $\varepsilon_{b,1}(\rho_0)$ shows that $\cal P$ now stays unimodal [Fig.~\ref{Fig:1}(c)], the increase/decrease of $r$ with $\varepsilon$ appears continuous, and the hysteresis is weak [inset of Fig.~\ref{Fig:1}(c)]. In other words, metastability progressively disappears when reducing $\rho_0$, showing how repulsion changes the nature of the transition.

% ------------------------------------------------------------------------------------------

\begin{figure}
  \includegraphics[width=\linewidth]{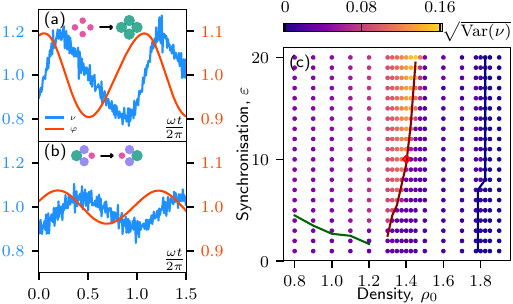}
  \caption{Packing fraction $\varphi$ and current $\nu$ oscillate in quadrature, with higher amplitude for (a)~globally synchronized particles than for (b)~dynamical patterns. Trajectories in (a) and (b) are sampled at the same point $(\varepsilon=10,\rho_0=1.4)$ of the phase diagram [red diamond in (c)].
  (c)~The current variance ${\rm Var}(\nu)$ changes abruptly close to the phase boundary $\varepsilon_{b,2}(\rho_0)$ [same line as in Fig.~\ref{Fig:1}(a)].
  }
	\label{Fig:2}
\end{figure}

\textit{Current statistics.}---Repulsion puts a strong constraint on phase dynamics whenever the packing fraction of large particles $\rho_0 \pi \sigma_0^2$ is comparable with close packing. In this regime, overlap between particles accelerates their contraction and decelerates their expansion. As a result, the packing fraction $\varphi$ oscillates in quadrature (phase shift of $\pi/2$ in natural units) with the scaled instantaneous phase velocity
\begin{equation}
	\nu = \frac{1}{N\omega}\sum_{j=1}^N\dot{\theta}_j .
\end{equation}
In the cycling ordered state, strong repulsion yields oscillations with high amplitude  [Fig.~\ref{Fig:2}(a)]. In the disordered state with patterns, phase shift between neighbours enables a smaller overlap and a reduced oscillation amplitude [Fig.~\ref{Fig:2}(b)]. We detect the corresponding transition in terms of the current variance $\text{Var}(\nu) = \langle\nu^2\rangle-\langle\nu\rangle^2$ [Fig.~\ref{Fig:2}(c)]. In the ordered state, $\text{Var}(\nu)$ increases steadily with $\rho_0$ until it abruptly drops. The location of this drop is close to the boundary $\varepsilon_{b,2}(\rho_0)$ [Fig.~\ref{Fig:1}(a)].

The averaged parameter $\langle\nu\rangle$ is close to $1$ throughout the phase diagram, except in the arrested ordered state ($\rho_0>1.8$) where $\nu$ vanishes~\cite{SM}. In that respect, deformation waves can be regarded as the collective strategy which, by introducing local phase shifts, enables to maintain high $\langle\nu\rangle$ even at high $\rho_0$. Importantly, synchronisation mitigates high phase shift, promoting bands with equal phase. Therefore, in contrast to previous works~\cite{Tjhung2016, Tjhung2017}, synchronisation here stands out as the essential ingredient for stabilizing patterns. In short, the transition at $\varepsilon_{b,2}(\rho_0)$ arises from a competition favoring either synchronisation at the cost of overlap, or smaller overlap at the cost of reduced synchronisation.

% ------------------------------------------------------------------------------------------

\textit{Hydrodynamics.}---The deformation waves in Figs.~\ref{Fig:0}(a-d) are reminiscent of the chemical waves in RDS~\cite{Turing1952, Kondo2010}. Although the microscopic details of PAM and RDS are quite different, it is tempting to draw analogies between their hydrodynamics. For simplicity, setting $\mu=0$, we neglect the role of repulsion in density fluctuations. In the phase dynamics, we assume that the interaction range is perfectly local: $\sum_j {\cal T}(a_{ij},\theta_i-\theta_j) \approx \pi\sigma_0^2\varepsilon \sum_j \sin(\theta_i-\theta_j) \,\delta({\bf r}_i - {\bf r}_j)$, and we treat  repulsion as $\sum_j \partial_{\theta_i} U(a_{ij}) \approx \pi\sigma_0^2(\partial_\varphi U) \sum_j (\partial_{\theta_j}\varphi) \,\delta({\bf r}_i - {\bf r}_j)$, where $\partial_\varphi U$ is again assumed constant. Using coarse-graining procedures~\cite{SM}, we obtain a closed dynamics for the local order parameter $A({\bf r},t) = \sum_j e^{{\rm i}\theta_j} \delta({\bf r} - {\bf r}_j(t))$ in powers of $A$ and its gradients:
\begin{equation}\label{hydro}
	\begin{aligned}
		\partial_t A &= \Big( \frac{\bar\varepsilon\rho_0}{2} - D_{\theta} + {\rm i}\omega + D\nabla^2\Big) A - \frac{\bar\varepsilon^2  A|A|^2}{4(2D_{\theta} - {\rm i}\omega)}
		\\
    &\quad - {\rm i} c A \bigg[ {\rm Re}(A) + \frac{\bar\varepsilon\lambda}{4} {\rm Im} \Big( \frac{A^2}{2D_{\theta} - {\rm i}\omega} \Big) \bigg] + \sqrt{\rho_0D_{\theta}} \,\Lambda ,
	\end{aligned}
\end{equation}
where $\bar\varepsilon = \pi\sigma_0^2\varepsilon$, and $c = \mu_\theta \lambda (\pi\sigma_0 \sigma(0)/L)^2 (\partial_\varphi U)$. The field $\Lambda$ is a Gaussian white noise with zero mean. This hydrodynamics is analogous to the complex Ginzburg-Landau equation describing a large class of RDS~\cite{Aranson2002}. The term proportional to $c$ in Eq.~\eqref{hydro}, which stems from the repulsive term $\partial_{\theta_i} U$ in Eq.~\eqref{eqn:MDb}, here breaks the gauge invariance $A\to A \,e^{{\rm i} \Phi}$ for any $\Phi$. This reveals how the coupling between repulsion and change in size at microscopic level affects the gauge symmetry of the hydrodynamic phase. Indeed, changing uniformly the phases of all particles does not leave the dynamics invariant due to particle overlap.

\begin{figure}[b]
  \includegraphics[width=\linewidth]{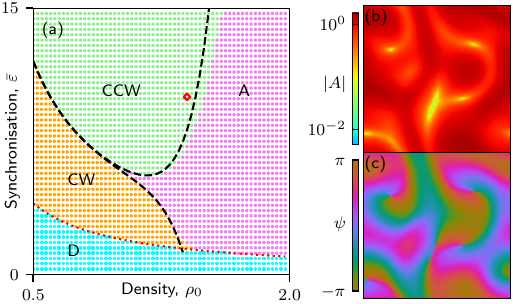}
  \caption{(a)~Phase diagram of homogeneous states in noiseless hydrodynamics: (D)~disorder, (A)~arrest, (CW)~clockwise, and (CCW)~counter-clockwise cycles. The red dotted line delineates order-disorder transition. The black dashed lines approximate the stability of arrest~\cite{SM}.
  (b-c)~In the presence of noise, steady states with motile defects appear for $(\bar\varepsilon,\rho_0)$ between cycling and arrest [diamond in (a)].
  }
	\label{Fig:3}
\end{figure}

In the noiseless dynamics, the homogeneous profile is always linearly stable. It can correspond to order $|A| > 0$ or disorder $|A| = 0$. The ordered state can be either arrested or cycling, respectively for constant and time-dependent $\psi$, where $A = |A|e^{{\rm i}\psi}$. The latter cycles either in the same direction as the microscopic model (clockwise, CW), or in the opposite direction (counter-clockwise, CCW). The CCW state emerges deep in the ordered state, while the derivation of Eq.~\eqref{hydro} relies on an expansion at small $A$, which should only be valid close to the disordered state. This explains why such a state has no counterpart in the microscopic model.

We obtain analytically the phase boundaries between these states [Fig.~\ref{Fig:3}(a)]. The order-disorder transition follows $\bar\varepsilon\rho_0 = 2D_\theta$, in line with mean-field arguments for the particle-based dynamics. The term proportional to $c$ in Eq.~\eqref{hydro} controls the existence of stationary solutions for $\psi$. This highlights how microscopic repulsion induces arrest in hydrodynamics. At high density, repulsion gets rapidly enhanced with $\rho_0$, so we take $c\propto\rho_0^n$ for $n>0$. The arrested state then arises at large $\rho_0$, and cycling occurs for intermediate density regimes between disorder and arrest, as expected. In short, our hydrodynamics reproduces the phase boundaries of the homogeneous states observed for particles [Fig.~\ref{Fig:1}(a)].

Due to non-linearities, homogeneous profiles are not the only stable solutions. In the presence of noise, between cycling and arrested states, the hydrodynamics now systematically reaches steady states with dynamical patterns [Figs.~\ref{Fig:3}(b-c)]. They are always associated with motile defects constantly forming and merging, reminiscent of defect turbulence [Fig.~\ref{Fig:0}(d)]. Interestingly, we do not observe waves with non-motile defects in contrast to microscopic simulations, suggesting that density fluctuations, neglected in our hydrodynamics, play an important role in stabilizing such structures.

% ------------------------------------------------------------------------------------------

\textit{Discussion.}---It is striking that PAM entails hydrodynamic patterns akin to RDS. Indeed, our model does not feature any reaction, and particle diffusion is strongly hampered in dense regimes of interest. Yet, it is insightful to regard individual pulsation as a chemical reaction, whose coordinate is the particle phase, with $\omega$ effectively driving cycles between isomers. Importantly, this drive is a monomolecular reaction. Consistently, hydrodynamic non-linearities all stem from particle interactions: setting $\bar\varepsilon=0=c$ yields a linear dynamics for $A$ [Eq.~\eqref{hydro}]. The interplay between synchronisation, repulsion and deformation is then akin to mechano-chemical coupling, which indeed regulates cell size in tissues~\cite{Howard2011, Recho2019}. From a broader perspective, our model can be regarded as a minimal, yet non-trivial description of a large class of systems involving synchronisation and mechanical pulsation.

Here, contraction is sustained by a microscopic periodic drive, in contrast with theories featuring contractile stress at the hydrodynamic level~\cite{Marchetti2013}. While the collective dynamics of deformable particles has recently attracted attention~\cite{Manning2023}, our model motivates one to explore further the role of mechanical pulsation. Importantly, the emergence of deformation waves  should be robust when considering even more complex interactions. For instance, by promoting the reference area and perimeter of deforming cells~\cite{Bi2015, Bi2016, Yeomans2019, Loewe2020, Lin2022} to pulsating variables with local synchronisation, the ensuing oscillations between elongated and rounded shapes is again a route to wave propagation.

Interestingly, our model reveals how density controls the crossover between wave types. In cardiac tissues, the emergence of spiral waves and turbulence signals fibrillation. Various strategies have been proposed to mitigate this while avoiding drastic measures~\cite{Luther2011, Karma2013, Fenton2022}. To this end, our model stands out as an appropriate platform in search of protocols inhibiting specific waves with local perturbation.

\acknowledgments{We acknowledge insightful discussions with Michael E. Cates, Luke K. Davis, Massimiliano Esposito, Robert L. Jack, Alessandro Manacorda, Xia-Qing Shi, Benjamin D. Simons, Julien Tailleur, and Fr\'ed\'eric van Wijland. Work funded by the Luxembourg National Research Fund (FNR), grant reference 14389168.}

\bibliography{References}

\begin{thebibliography}{57}
\bibitem [{mov()}]{movies}%
  \BibitemOpen
  \href@noop {} {}\bibinfo {note} {See movies of numerical simulations at [URL
  will be inserted by publisher] corresponding to snapshots shown in
  Fig.~\ref{Fig:0}.}\BibitemShut {Stop}%
\end{thebibliography}

\end{document}